# Reconstruction of the reverberation theory in a diffuse sound field by using reflection orders

Toshiki Hanyu [1*]

[1] *Nihon University, Junior College, Dept. of Architecture and Living Design, 7-24-1 Narashinodai, Funabashi, Chiba, 274-8501 Japan*

**Abstract:** Room acoustics is mainly based on the reverberation theories of Saine and Eyring. In Sabine's theory however, the reverberation time does not reach zero, even if the condition of absolute absorption is fulfilled. Eyring revised reverberation theory to resolve this contradiction. However, Eyring's theory has an inconsistency between the formulations of the steady-state and decay processes. Therefore, the author revised Sabine's theory, taking a different approach from that of Eyring. This revised theory was constructed by introducing the concept of "reverberation of a direct sound." In this study, a new mathematical model of reverberation using reflection orders is proposed. This is a reconstruction of the author's revised theory. The new model includes the temporal energy distribution in each reflection order and uses the concept of "reverberation of a direct sound" for the entire reverberation process. It shows that the concept is also essential for the reflected sounds. In addition, the reverberation decay agrees with the revised theory previously proposed by the author. Overall, the new model showed good agreement with the simulation results.

**Keywords:** Reverberation theory, Reflection orders, Sabine, Eyring, Revised theory

## 1. INTRODUCTION

The reverberation theories of Sabine [1] and Eyring [2] are still the cornerstones of room acoustics. These theories assume the diffuse sound field. Under this assumption, important physical quantities such as the average sound pressure level (sound energy density) and the reverberation time (sound energy decay) can be derived from these theories. However, Sabine's theory is considered to contain a contradiction in that the reverberation time does not become zero even when the sound field is in a state of complete absorption. Eyring revised Sabine's theory to resolve this contradiction [2].

As real sound fields are not diffuse sound fields, many attempts have been made to adapt the reverberation theory to non-diffuse sound fields from different angles, such as the concept of directional reverberation [3,4], non-uniform absorption [5-7] and the consideration of sound scattering [8-10]. In contrast, there have been almost no attempts to revise the diffuse-field theories of Sabine and Eyring.

The author has made it clear that Eyring's theory has an inconsistency in the formulations between the steady-state and decay processes [11]. That is, Sabine's and Eyring's theories do not form an integrated and consistent theoretical system. Therefore, to construct a consistent theoretical system for room acoustics, the author has revised Sabine's reverberation theory [11], taking a different approach from that of Eyring. This revised theory was constructed by introducing the concept of "reverberation of a direct sound." In the revised theory, the concept of "reverberation of a direct sound" is essential not only for the direct sound, but also for the entire reverberation process.

The objective of this study was to describe the revised theory mathematically in a different way and to reconstruct it in such a way that the temporal energy distribution in each reflection order is accounted for [12].

Polack proposed a mathematical model based on a Poisson distribution for the temporal distribution of the reflected sounds, a generalization of Sabine's formula [13]. However, similar to Sabine's formula, it was believed that this model was not valid for a room with complete absorption. The present author has attempted to express mathematically the time distribution of diffuse sound in a room for each reflection order using a normal distribution [14]. However, the trial was not completed because the idea of "reverberation of a direct sound" did not exist at that time.

In this paper, it is shown that the new mathematical model using reflection orders is basically consistent with the revised theory already proposed by the author [11], and that the concept of "reverberation of a direct sound" is also essential for the entire reverberation process.

---

* hanyu.toshiki@nihon-u.ac.jp

## 2. EYRING'S THEORY IN TERMS OF REFLECTION ORDERS

In this study, a diffuse sound field was assumed. To fulfill this assumption, a room must not have extreme dimensions or an uneven absorption distribution. It is also assumed that the reflections from the wall are completely diffused and follow Lambert's cosine law.

It is assumed that a sound source with sound power $W$ emits sound for a short time $\bar{\ell}/c$ in a room with volume $V$, total surface area $S$, and the average absorption coefficient $\bar{\alpha}$, where $\bar{\ell}$ is the mean free path $4V/S$ [15-18], and $c$ is the speed of sound. Under these conditions, the sound energy density of the $n$th reflection order $E_n$ can be expressed by Eq. (1), based on Eyring's theory.

$$E_n = \frac{W}{V} \cdot \frac{\bar{\ell}}{c}(1-\bar{\alpha})^n = \frac{4W}{cS} \cdot (1-\bar{\alpha})^n, \quad (1)$$

where $n$ is the reflection order and is an integer.

According to Eq. (1), the reflected sound energy is not generated at all when $\bar{\alpha} = 1$. This resolves Sabine's contradictions. When $n=0$, Eq. (1) can be considered as direct sound energy. Thus, when $n=0$ and $\bar{\alpha} = 1$, Eq. (1) leads to another contradiction: the direct sound energy disappears completely, although the direct sound is not related to the wall absorption. Therefore, only the reflected sound with the condition $1 \leq n$ is considered here.

The energy density of the reflected sound in the steady state, $E_r$, can be calculated using Eq. (2). Based on Eq. (2), $E_r$ becomes $4W/cR$ as the sum of all reflection orders, where $R$ is the room constant and $R = S\bar{\alpha}/(1-\bar{\alpha})$.

$$E_r = \sum_{n=1}^{\infty} E_n = \frac{4W}{cS} \sum_{n=1}^{\infty}(1-\bar{\alpha})^n$$
$$= \frac{4W}{cS} \frac{1-\bar{\alpha}}{\bar{\alpha}} = \frac{4W}{cR} \quad (2)$$

The sound energy density of steady state $E_0$ including the direct sound, can be expressed using Eq. (3).

$$E_0 = \frac{4W}{cS}\left[1 + \sum_{n=1}^{\infty}(1-\bar{\alpha})^n\right] = \frac{4W}{cS\bar{\alpha}}. \quad (3)$$

Eq. (1) represents the energy per reflection order but does not account for time $t$ and does not represent the reverberation decay. In Eyring's theory, the reverberation decay can be obtained as a function of time $t$ as shown in Eq. (4) by replacing $n$ in $(1-\bar{\alpha})^n$ of Eq. (1) with the average reflection order $m = cSt/4V$ until time $t$.

$$(1-\bar{\alpha})^m = \exp\left[\frac{cS\ln(1-\bar{\alpha})}{4V}t\right]. \quad (4)$$

where $1 \leq m$, $\bar{\ell}/c \leq t$ and $m$ is a real number. Based on Eqs. (3) and (4), the reverberation decay from the steady-state in Eyring's theory can be expressed using Eq. (5), as follows:

$$E(t) = \frac{4W}{cS\bar{\alpha}}(1-\bar{\alpha})^m = \frac{4W}{cS\bar{\alpha}}\exp\left[\frac{cS\ln(1-\bar{\alpha})}{4V}t\right]. \quad (5)$$

Eq. (5) is a function of the average reflection order $m$ or time $t$. In contrast, Eq. (5), contains no information about the reflection order $n$, which is an integer. In other words, Eyring's theory can express the energy density for each reflection order $n$ as in Eq. (1), but it does not include information about the temporal energy distribution in each reflection order $n$.

## 3. OUTLINE OF THE REVISED REVERBERATION THEORY

This chapter provides an overview of the revised theory previously proposed by the author [11].

In the revised theory, reverberation is defined as "a decay of the average sound energy density throughout the space, which includes both direct sound and reflected sounds." Moreover, it was assumed that a perfect diffusion state is maintained both in a steady state and throughout the reverberation process.

It is assumed that a sound source with sound power $W$ in a room with volume $V$ emits sound for an extremely short time $\Delta t$ and that the sound energy density $E$ decays exponentially with a decay parameter $\lambda$. This can be expressed using Eq. (6). This equation expresses the average impulse response in the room.

$$\Delta E(t) = \frac{W\Delta t}{V}\exp(-\lambda t). \quad (6)$$

Eq. (7) is obtained by setting $\Delta t \to 0$.

$$\frac{dE(t)}{dt} = \frac{W}{V}\exp(-\lambda t). \quad (7)$$

The Schroeder integration [19] in Eq. (7) results in the following equation, which expresses the sound energy decay from the steady state of the sound field:

$$E(t) = \frac{W}{V}\int_t^{\infty} \exp(-\lambda\tau)\,d\tau = \frac{W}{V}\frac{1}{\lambda}\exp(-\lambda t). \quad (8)$$

From Eq. (8), the sound energy density $E_0$ in the steady state and the energy decay can be defined as $W/V\lambda$ and $\exp(-\lambda t)$ respectively. In Sabine's theory, $\lambda$ becomes $cS\bar{\alpha}/4V$ and the reverberation decay from the steady state can be expressed by Eq. (9).

$$E(t) = \frac{W}{V}\frac{4V}{cS\bar{\alpha}}\exp\left(-\frac{cS\bar{\alpha}}{4V}t\right). \quad (9)$$

Fig. 1 shows a conceptual diagram of reverberation decay based on Eq. (7). According to Sabine's theory, the reverberation decay in Eq. (7) becomes $(W/V)\exp(-ct/\bar{\ell})$ for $\bar{\alpha}=1$, where $\bar{\ell} = 4V/S$. This decay has been considered as a contradiction in Sabine's theory because reflected sound is not generated at all when $\bar{\alpha}=1$.

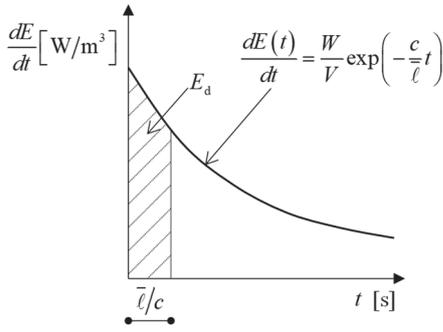

**Fig. 1** Conceptual diagram of the energy density of direct sound [11].

In the revised theory, this decay is construed as a reverberation of direct sound. The sound energy of the direct sound does not disappear until it reaches the wall surfaces, even when $\bar{\alpha}=1$. Thus, if the sound source stops at a steady state, the sound energy of the direct sound remains in space for some time. This is the concept of "reverberation of direct sound." According to this concept, the reverberation time does not necessarily reach zero, contrary to Eyring's theory. However, it is incorrect to assume that the direct sound energy persists far beyond the time $t = \bar{\ell}/c$, as in Sabine's theory.

To resolve the problem of the reverberation of direct sound in Sabine's theory, the maximum time that direct sound can remain $t_{d\_max}$ should be examined. Firstly, the free path of direct sound $\ell_d$ is defined as the distance the sound travels from a sound source to the wall. If the maximum length of $\ell_d$ is defined as $\ell_{d\_max}$, $t_{d\_max}$ can be expressed as $\ell_{d\_max}/c$. However, $\ell_{d\_max}$ cannot be determined without a specific room shape. Therefore, based on the stochastic concept, the expected value $\overline{\ell_{d\_max}}$ must be used instead of $\ell_{d\_max}$. In this case, the expected values of $t_{d\_max}$, $\overline{t_{d\_max}}$, can be derived as $\overline{\ell_{d\_max}}/c$. $\overline{\ell_{d\_max}}$ can be deduced by assuming an extreme condition where there are numerous sound sources on all wall surfaces. Under this condition, it can be assumed that $\overline{\ell_{d\_max}}$ is equal to the mean free path $\bar{\ell}$. Therefore, the direct sound is limited to the time $\overline{t_{d\_max}} = \bar{\ell}/c$ on average in the revised theory. On this basis, the energy density of the direct sound $E_d$ can be calculated using Eq. (10).

$$E_d = \frac{W}{V}\int_0^{\frac{\bar{\ell}}{c}}\exp\left(-\frac{c}{\bar{\ell}}\tau\right)d\tau = \frac{W}{V}\frac{\bar{\ell}}{c}(1-e^{-1}). \quad (10)$$

The energy density of the direct sound $E_d$ becomes $W\bar{\ell}/cV$ in the theories of Sabine and Eyring, but $W\bar{\ell}(1-e^{-1})/cV$ in the revised theory. Eq. (10) contains only direct sound, as Eq. (10) can be derived from Eq. (7) with the conditions $\lambda = cS\bar{\alpha}/4V$ and $\bar{\alpha} = 1$. Therefore, $E_d$ can be calculated by setting $\bar{\ell}/c$ as the integral interval.

$E_0$ is redefined in the revised theory, according to Eyring's method using $E_d$ calculated by Eq. (10). Based on the above discussion, $E_0$ can be calculated using Eq. (11).

$$\begin{aligned}E_0 &= \frac{W}{V}\frac{\bar{\ell}}{c}\left[(1-e^{-1}) + \sum_{n=1}^{\infty}(1-\bar{\alpha})^n\right] \\ &= \frac{W}{V}\frac{4V(1-e^{-1}\bar{\alpha})}{cS\bar{\alpha}}\end{aligned} \quad (11)$$

Comparing Eq. (11) to (8), parameter $\lambda$ in the revised theory can be derived as $cS\bar{\alpha}/4V(1-e^{-1}\bar{\alpha})$. Using this parameter $\lambda$, the reverberation decay from the steady state can be expressed using Eq. (12).

$$\begin{aligned}E(t) &= \frac{W}{V}\frac{4V(1-e^{-1}\bar{\alpha})}{cS\bar{\alpha}}\exp\left[-\frac{cS\bar{\alpha}}{4V(1-e^{-1}\bar{\alpha})}t\right] \\ &= \frac{4W}{c}\left(\frac{1}{R}+\frac{1-e^{-1}}{S}\right)\exp\left[-\left(\frac{1}{R}+\frac{1-e^{-1}}{S}\right)^{-1}\frac{ct}{4V}\right]\end{aligned} \quad (12)$$

When the mean free path of the direct sound is $\overline{\ell_d}$, the energy density of the direct sound $E_d$ can be expressed as $(W/V)(\overline{\ell_d}/c)$. Comparing this to Eq. (10), the mean free path of direct sound $\overline{\ell_d}$ can be expressed by Eq. (13).

$$\overline{\ell_d} = \bar{\ell}(1-e^{-1}) = \frac{4V}{S}(1-e^{-1}). \quad (13)$$

Eq. (12) is a mixture of the reverberation decays of the direct and reflected sounds. Substituting $t=0$ into Eq. (12) gives the sound energy density of the reflected sound $4W/cR$ and that of the direct sound $4W(1-e^{-1})/cS$ at the steady state. If $\bar{\alpha} = 1$ is used, the reverberation decay of the direct sound $E_d(t)$ can be expressed by Eq. (14).

$$E_d(t) = \frac{4W(1-e^{-1})}{cS}\exp\left[-\left(\frac{1-e^{-1}}{S}\right)^{-1}\frac{ct}{4V}\right]. \quad (14)$$

Subtracting Eq. (14) from Eq. (12) gives the reverberation decay curve $E_r(t)$ for the reflected sound only in the revised theory, as shown in Eq. (15). Substituting $t=0$ into Eq. (15) gives the sound energy density of the reflected sound, $4W/cR$. If $\bar{\alpha} = 1$ is used, $E_r(t) = 0$, which confirms that there is no reverberation of the reflected sound.

$$E_r(t) = E(t) - E_d(t)$$
$$= \frac{4W}{c}\left(\frac{1}{R} + \frac{1-e^{-1}}{S}\right)\exp\left[-\left(\frac{1}{R} + \frac{1-e^{-1}}{S}\right)^{-1}\frac{ct}{4V}\right] \quad (15)$$
$$- \frac{4W(1-e^{-1})}{cS}\exp\left[-\left(\frac{1-e^{-1}}{S}\right)^{-1}\frac{ct}{4V}\right]$$

## 4. RECONSTRUCTION OF REVERBERATION THEORY USING REFLECTION ORDERS

The proposed revised theory is described mathematically in a different way. The revised theory was reconstructed by revising Eyring's reverberation theory to account for temporal energy distributions in each order of reflection.

This new mathematical model uses the concept of "reverberation of a direct sound" to also describes the reverberation of reflected sounds. This shows that the concept of "reverberation of a direct sound" is also essential for the entire reverberation process.

### 4.1. Theoretical framework

The first step was to develop a theoretical framework for the temporal distribution of reflected sound energy for each reflection order. Then, a "reverberation by reflected sound" should be obtained by adding up all the reflection orders.

A conceptual diagram of the probability density function $P_n(t)$ of the temporal distribution of the $n$th order reflected sound is shown in Fig. 2, where $P_n(t)$ is a probability density function whose integration is 1.

$$\int_0^\infty P_n(t)dt = 1. \quad (16)$$

It is assumed that the mean value of $P_n(t)$ increases with $\bar{\ell}/c$ for each reflection order. The addition of $P_n(t)$ for all reflection orders under these conditions results in a constant value $c/\bar{\ell}$ at $0 \ll t$.

$$\sum_{n=1}^{\infty} P_n(t) = \frac{c}{\bar{\ell}} \quad \text{const.}(0 \ll t). \quad (17)$$

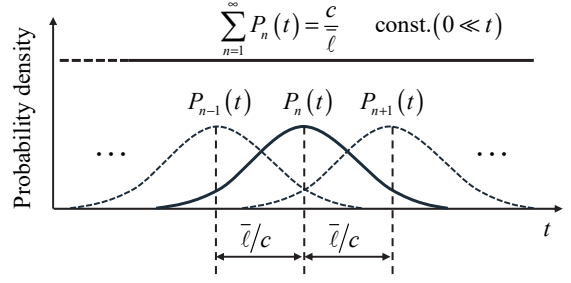

**Fig. 2** Concept of the probability density of the time distribution of the $n$th order reflected sound.

Fig. 2 is converted into a conceptual diagram of the time variation of the proportion of $n$th order reflected sound, as shown in Fig. 3. In Fig. 3, $P_n(t)$ is multiplied by $\bar{\ell}/c$ to set the probability 1 at $0 \ll t$ when $P_n(t)$ for all reflection orders are added, as shown in Eq. (18). $\bar{\ell}/c \times P_n(t)$ represents the proportion of $n$th order reflected sound at time $t$.

$$\sum_{n=1}^{\infty} \frac{\bar{\ell}}{c} P_n(t) = 1 \quad \text{const.}(0 \ll t). \quad (18)$$

In Eq. (18), the probability 1, independent of time, corresponds to the fact that, if sound absorption is not considered, the sound energy emitted in the room persists with probability 1, with repeated reflections and increasing reflection orders.

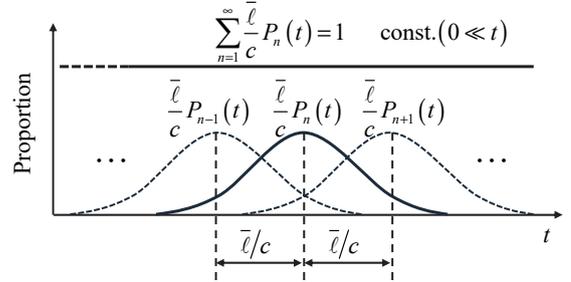

**Fig. 3** Concept of the time variation of the proportion of the $n$th order reflected sound.

Based on the concept shown in Fig. 3, and assuming a diffuse sound field, it is assumed that the sound is emitted for a short time $\Delta t$ from a source with sound power $W$ in a room with volume $V$. In this case, the temporal distribution of the reflected sound energy density for the reflection order $n$ can be expressed as shown in Eq. (19). This can be interpreted as the extraction of only the $n$th order reflected sound from the average impulse response defined in Eq. (6). In other words, Eq. (19) represents the average impulse response of the $n$th order reflection.

$$\Delta E_n(t) = \frac{W\Delta t}{V} \cdot (1-\bar{\alpha})^n \cdot \frac{\bar{\ell}}{c} P_n(t). \quad (19)$$

Eq. (20) is obtained by setting $\Delta t \to 0$.

$$\frac{dE_n(t)}{dt} = \frac{W}{V} \cdot (1-\bar{\alpha})^n \cdot \frac{\bar{\ell}}{c} P_n(t). \qquad (20)$$

As in Eq. (21), the Schröder integration of $P_n(t)$ can be defined here as $S_n(t)$. A conceptual diagram of $S_n(t)$ is shown in Fig. 4. From Eq. (16) it can be understood that $S_n(0) = 1$.

$$\int_t^\infty P_n(\tau)d\tau = S_n(t). \qquad (21)$$

Using $S_n(t)$ and the Schröder integration of Eq. (20), the energy decay $E_n(t)$ of only the $n$th order reflected sound from the steady state is obtained, as shown in Eq. (22).

$$\begin{aligned} E_n(t) &= \frac{W}{V} \cdot (1-\bar{\alpha})^n \cdot \frac{\bar{\ell}}{c} \int_t^\infty P_n(\tau)d\tau \\ &= \frac{W}{V} \cdot (1-\bar{\alpha})^n \cdot \frac{\bar{\ell}}{c} S_n(t) \\ &= \frac{4W}{cS} \cdot (1-\bar{\alpha})^n \cdot S_n(t) \end{aligned} \qquad (22)$$

Finally, by adding the energy decay $E_n(t)$ of the $n$th order reflected sound over all reflection orders, the reverberation decay $E_r(t)$ due to the reflected sound from the steady state is obtained as shown in Eq. (23).

$$\begin{aligned} E_r(t) &= \sum_{n=1}^\infty E_n(t) \\ &= \frac{4W}{cS} \sum_{n=1}^\infty \left[ (1-\bar{\alpha})^n \cdot S_n(t) \right] \end{aligned} \qquad (23)$$

In Eq. (23) the reflection order $n$ is preserved as an integer. Therefore, the total reflected sound is described as the sum of all reflection orders, as shown in Eyring's equation (2). The difference between Eq. (2) and Eq. (23) is that Eq. (23) contains information about the decay in each reflection order. This allows the reverberation decay to be decomposed and calculated for each reflection order.

The onset of decay, $E_r(0)$, i.e., the sound energy density due to the reflected sound at the steady state, can be expressed as shown in Eq. (24), where $S_n(0) = 1$.

$$\begin{aligned} E_r(0) &= \frac{4W}{cS} \sum_{n=1}^\infty \left[ (1-\bar{\alpha})^n \cdot S_n(0) \right] \\ &= \frac{4W}{cS} \sum_{n=1}^\infty (1-\bar{\alpha})^n = \frac{4W}{cR} \end{aligned} \qquad (24)$$

Eq. (24) is identical to Eq. (2) from Eyring's theory. In other words, Eq. (23) also shows that the sound energy density of the reflected sound in the steady state is $4W/cR$, as in Eyring's theory.

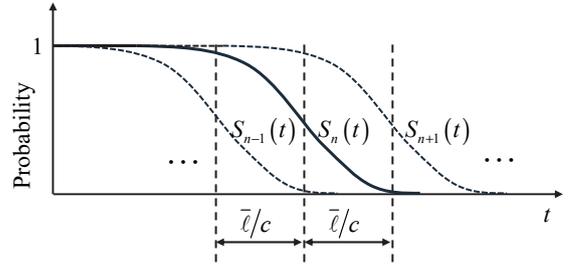

**Fig. 4** Schroeder integration of the probability density of the $n$th order reflected sound.

### 4.2. New model using reflection orders

In this section, the revised reverberation theory is reconstructed as a new mathematical model based on the theoretical framework that accounts for the reflection orders described in the previous section. This comes down to how the probability density $P_n(t)$ in Eq. (16) is set.

Both the revised theory and new model are based on the stochastic concept and describe the expected value of the effect of a particular sound source under the condition of perfect diffusion, both in the steady state and during the entire reverberation process. This is equivalent to describing the average effect of numerous point sound sources that are uniformly distributed throughout the room.

The minimum sound energy that could no longer be separated was assumed to be a sound particle. Each point source radiated numerous sound particles in each direction. This allows a point sound source to be simulated.

A particle travels at the speed of sound in a room. When a particle hits a wall surface, it is assumed that the reflections from the wall are completely diffused, according to Lambert's cosine law. Based on these assumptions, after repeated multiple reflections, the probability distribution of the $n$th order reflected sound on the time axis converges to a normal distribution according to the "law of large numbers" and the "central limit theorem."

For the sake of simplicity, a mathematical model has been constructed that assumes a normal distribution for all reflection orders. A random variable $X$ is defined that follows a normal distribution $N(\bar{\ell}/c, \{\overline{\ell_d}/c\}^2)$ where the average value $\mu$ is the mean free path $\bar{\ell}/c$ and the standard deviation $\sigma$ is $\overline{\ell_d}/c$ as shown in Eq. (25), and Fig.5. The random variable $X$ represents the probability of the presence of the reflected sound at time $t$. $\overline{\ell_d}$ is the mean free path of the direct sound, expressed by Eq. (13). $\overline{\ell_d}$ can also be interpreted as the average distance that the sound energy in a diffuse sound field travels before it hits the wall surface. If the perfect diffusion state is maintained throughout the reverberation process, the concept of $\overline{\ell_d}$ applies not only to the direct sound but also to the reflected sound throughout the reverberation decay process.

$$X \sim N\left(\frac{\overline{\ell}}{c}, \left\{\frac{\overline{\ell_d}}{c}\right\}^2\right). \tag{25}$$

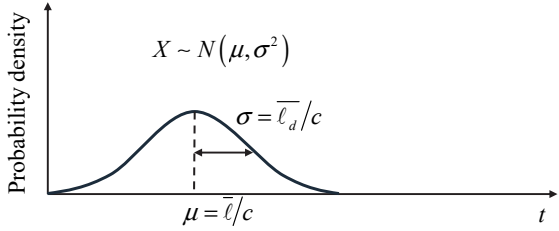

**Fig. 5** Stochastic variable $X$ with normal distribution $N(\mu, \sigma^2)$.

Let $X_n$ be a random variable that follows the probability density of the temporal distribution of the $n$th order reflected sound. As the sound particles are scattered at the wall surface with a probability of 1, each reflection can be regarded as an independent event. Therefore, as expressed in Eq. (26), adding the random variable $X$ $n$ times results in the random variable $X_n$, which represents the $n$th order reflections.

$$X_n = \sum_{i=1}^{n} X + b \sim N(n\mu + b, n\sigma^2). \tag{26}$$

The constant $b$ represents the overall time shift in the time distribution of the reflected sound. Although there is still room for further investigation, it was assumed that the time $\overline{\ell_d}/2c$ is the time required for the sound to travel half the distance of the mean free path of the direct sound. Based on the above, the probability density function $f_n(t)$ of the random variable $X_n$ is defined in Eq. (27).

$$\begin{aligned}
f_n(t) &= N(n\mu + b, n\sigma^2) \\
&= N\left(n\frac{\overline{\ell}}{c} + \frac{\overline{\ell_d}}{2c}, n\left(\frac{\overline{\ell_d}}{c}\right)^2\right) \\
&= \frac{1}{\sqrt{2\pi n\left(\frac{\overline{\ell_d}}{c}\right)^2}} \exp\left[-\frac{\left\{t - \left(n\frac{\overline{\ell}}{c} + \frac{\overline{\ell_d}}{2c}\right)\right\}^2}{2n\left(\frac{\overline{\ell_d}}{c}\right)^2}\right]
\end{aligned} \tag{27}$$

Eq. (27) is a probability density function that satisfies Eq. (28).

$$\int_{-\infty}^{\infty} f_n(t) dt = 1. \tag{28}$$

However, Eq. (27) shows that $f_n(t) \neq 0$ when $t \leq 0$, a contradiction stating that the probability of the presence of reflected sound is not zero even when $t \leq 0$. To solve this problem, the decay term of direct sound in Eq. (14) is defined as $P_d(t)$ as expressed in Eq. (29).

$$P_d(t) = \begin{cases} \exp\left[-\left(\frac{1-e^{-1}}{S}\right)^{-1} \frac{ct}{4V}\right] & \text{if } 0 \leq t \\ 0 & \text{if } t < 0 \end{cases}. \tag{29}$$

As the temporal distribution of the reflected sound, $f_n(t)$ can be multiplied by $\{1 - P_d(t)\}$ to satisfy $\{1 - P_d(t)\}f_n(t) = 0$ when $t \leq 0$. However, if the integration interval is set to 0 to $\infty$ as in Eq. (16), the integral of $\{1 - P_d(t)\}f_n(t)$ is less than 1, $K_n$, as shown in Eq. (30). Therefore, the substitution of $\{1 - P_d(t)\}f_n(t)$ into $P_n(t)$ does not fulfill Eq. (16).

$$\int_0^{\infty} \{1 - P_d(t)\} f_n(t) dt = K_n < 1. \tag{30}$$

The probability density function $P_n(t)$ of the temporal distribution of the $n$th order reflected sound is then defined as Eq. (31).

$$P_n(t) = \frac{\{1 - P_d(t)\} f_n(t)}{K_n}. \tag{31}$$

$P_n(t)$ defined by Eq. (31) satisfies Eq. (16). Substituting this Eq. (31) into Eq. (22) as the probability density function $P_n(t)$ of the temporal distribution of the $n$th order reflected sound, the energy decay $E_n(t)$ of the $n$th order reflected sound in the steady state can be obtained. The reverberation due to the reflected sound $E_r(t)$ can also be calculated using Eq. (23).

## 5. COMPARISON OF THEORIES WITH COMPUTER SIMULATIONS

### 5.1. Simulation method

To verify the new model using reflection orders, simulations were performed using the sound ray tracing method. The energy decay of the reflected sound $E_r(t)$ and the probability density function of $n$th order reflected sound at time $t$, $P_n(t)$, and the Schröder integration of $P_n(t)$ namely $S_n(t)$ were calculated from the simulations and compared with the values calculated using the new model with reflection orders.

Traveling sound particles (representing the sound energy) were used as sound rays to simulate the propagation of sound energy. The energy of the sound particle was multiplied by $(1-\alpha)$ corresponding to the absorption coefficient $\alpha$ of the wall surface at each reflection. When each sound particle is reflected from the wall, it is diffusely reflected according to Lambert's cosine law.

A cubic room with a side length of 10 m and a rectangular room with dimensions of 15 × 10 × 5 m were used for the simulation. The absorption coefficients were set from 0.2 to 0.5 in steps of 0.1 and were the same for all walls.

The new model was constructed based on the average impulse response of the $n$th order reflection described in Section 4.1. Thus, the average impulse response of the $n$th order reflection is simulated first, and the reverberation decay is then determined by Schroeder integration of the average impulse response. Many point sources must be uniformly distributed to simulate a perfect diffusion state as the initial state of the sound field. Due to limited computer resources and time, it was not possible to use these conditions. To simulate numerous point sources, $10^8$ sound particles were randomly distributed throughout the room. The direction of propagation of each particle was determined randomly using uniform spherical random numbers. This corresponds statistically to a uniform distribution of an infinite number of point sound sources in a room. The total sound power $W$ of all the particles was set to 1.0 [watt]. The time step $\Delta t$ was $0.01/c$ because the travel of sound particles is calculated every 0.01 m. Because of the simulated numerous point sound sources and the diffuse reflections according to Lambert's cosine law in the steady state and reverberation process, the diffusion state can be approximately maintained.

From the total energy of the particles of each reflection order $n$ at each time step $\Delta t$, the change in the sound energy density $\Delta E_n$ over time, namely the average impulse response of the $n$th order reflection, was calculated. The reverberation decay curve of each reflection order $E_n(t)$ was then calculated by Schroeder integration of the average impulse response of the $n$th order reflection. The calculated $E_n(t)$ is the energy decay from steady state. Finally, by adding the energy decay $E_n(t)$ of the $n$th order reflected sound over all reflection orders, the reverberation decay $E_r(t)$ owing to the reflected sound from the steady state is obtained using Eq. (23). Simultaneously, the total number of particles of each reflection order $n$ was recorded at each time step $\Delta t$. Based on this information, $P_n(t)$, and $S_n(t)$ were calculated.

## 5.2. Results and discussion
### 5.2.1 Comparison of the theories to the simulation

Fig. 6 shows a comparison of the theoretical reverberation decay curves with the simulation results for a cubic room with a side length of 10 m and a rectangular room with dimensions of 15 ×10 ×5 m. The numbers on the decay curves represent the average absorption coefficients. The legend 'Revised' in the upper left graphs of the two rooms means the result using Eq. (15) for the revised theory. The results of the new model using the reflection orders according to Eq. (23) are shown in the upper-right graphs.

The results of the revised theory and the new model agree well with the simulation results. This also means that the results of the revised theory and the new model are almost identical. These results show that the new model using reflection orders is as valid as the previously proposed revised theory. The reverberation decay curves based on Sabine's theory showed a slower decay and an overestimation of reverberation times compared to the simulation results. In contrast, Eyring's results showed a faster decay and underestimated reverberation times compared to the simulation results. These discrepancies increase as the absorption coefficient increases. This means that the differences between the new models and existing theories increase with increasing absorption coefficients.

In the new model, $\overline{\ell_d}/c$ is used for the standard deviation $\sigma$ of the normal distribution for the random variable $X$. In fact, $\sigma$ influenced the decay curves. Fig. 7 shows decay curves calculated using different standard deviations $\sigma$, $\bar{\ell}/c$ and $\bar{\ell}/4c$. The smaller $\sigma$, the faster the decay. As shown in Fig. 6, the decay curves of the new model using $\overline{\ell_d}/c$ as $\sigma$, achieved the best agreement with the revised theory and the simulation results.

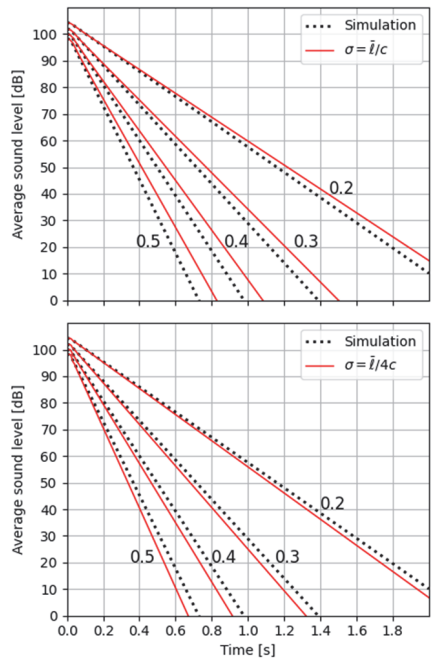

Cubic room with a side length of 10 m
**Fig. 7** Effects on the standard deviation $\sigma$ of the normal distribution for the random variable $X$.

### 5.2.2 Detailed comparison in terms of reflection orders

Figs. 8 and 9 show the results of the new model using reflection orders and ray-tracing simulations, respectively. From top to bottom, both figures show the reverberation decay curves using the reflection orders, the probability density function of each reflection order, and the Schröder integral of the probability density function. The top graphs of Figs. 8 and 9 show the decay curves for each reflection order and the overall decay curve. The overall decay curves of the new model and the simulation were obtained by the energetic summation of the decay curves for reflections of all orders. The decay curve calculated using the revised theory is also shown for comparison.

Looking first at the reverberation decay curves in Fig. 8, it is confirmed that the new model and the revised theory are almost identical.

Looking at the reverberation decay curves in Fig. 9, the results of the revised theory and the simulation are almost identical. This indicates that the revised theory is valid, as examined in our previous study [11]. The results in Figs. 8 and 9 show that the new model using the reflection orders examined in this study is as valid as the previously proposed revised theory.

Next, the probability density function for each reflection order is considered in detail, as shown in Figs 8 and 9. The probability density function for each reflection order shown in Fig. 8 was calculated using Eq. (25)–(31), and that shown in Fig. 9 was calculated indirectly from the information on the number of particles per reflection order of the ray-tracing simulation. This shows that the probability density function of each reflection order approaches a normal distribution as the reflection order $n$ increases. Furthermore, as the reflection order $n$ increases, the standard deviation of the normal distribution increases, and the temporal distribution gradually widens. The results of the new model in Fig. 8 show the same tendency as the simulation results in Fig. 9, except for the results for the lower reflection orders.

Next, the Schröder integral $S_n(t)$ of $P_n(t)$ is considered in detail. The results shown in Figs. 8 and 9 agree well with each other. $S_n(t)$ represents the reverberation decay for each reflection order, and the decay becomes gradually slower as the reflection order $n$ increases. According to Eq. (23), the reverberation decay curve for all reflected sounds can be obtained by multiplying the decay $S_n(t)$ for each reflection order $n$ by $4W(1-\bar{\alpha})^n/cS$ and then adding up all the reflection orders. The overall decay curves in the upper graphs of Figs. 8 and 9 show the results of this calculation procedure.

### 5.2.3 Difference between the new model and Eyring's theory

Here, an attempt was made to express Eyring's theory in terms of the probability density function of each reflection order, similar to the new model, and explain the differences between the new model and Eyring's theory. Fig.10 shows the calculation results using the probability density function with the identical standard deviation $\sigma = \overline{\ell_d}/c$ for all of the random variables $X_n$. For the calculation, $X_n \sim N(n\mu + b, \sigma^2)$ was used instead of $X_n \sim N(n\mu + b, n\sigma^2)$ in Eq. (26).

In the middle graph, the standard deviations of the probability density functions are constant, except for the lower orders. This differs from the new model and the simulation results shown in Figs 8 and 9. The decay curves for each reflection order and the overall decay curve, $E_r(t)$, are shown in the top graph, together with the decay curves calculated by Eyring's theory and the revised theory for comparison. The overall decay $E_r(t)$ determined with the constant $\sigma$ is almost identical to Eyring's decay curve. However, the overall decay $E_r(t)$ and Eyring's decay do not agree with the results of the revised theory and the simulation. This means that the expansion of the standard deviation with reflection orders, $X_n \sim N(n\mu + b, n\sigma^2)$, is essential and important for the mathematical model to represent the reverberation in a diffuse sound field. Eyring however, lacks this perspective.

### 6. CONCLUSIONS

In this study, a new mathematical model for reverberation using the reflection orders has been proposed. This also means a reconstruction of the revised theory previously proposed by the author to include the temporal energy distributions in each order of reflection. The new model also uses the concept of "reverberation of a direct sound" for the entire reverberation process.

The reverberation decay calculated with the new model agrees with the proposed revised theory. The new model showed good agreement with the simulation results. From these results, it can be concluded that the new mathematical model of reverberation essentially agrees with the revised theory already proposed by the author [11], and that the concept of "reverberation of a direct sound" is also essential for the reflected sounds.

Finally, the differences between the proposed model and Eyring's theory were also examined. As a result, it is clarified that expanding the standard deviation with reflection orders in the probability density functions of the reflected sounds is essential to express reverberation in a diffuse sound field, and that Eyring's theory lacks this aspect.

In the future, it will be necessary to verify the revised theory and the new model under various sound field conditions. Future work will also investigate the applicability of these new models to design and engineering in architectural acoustics.

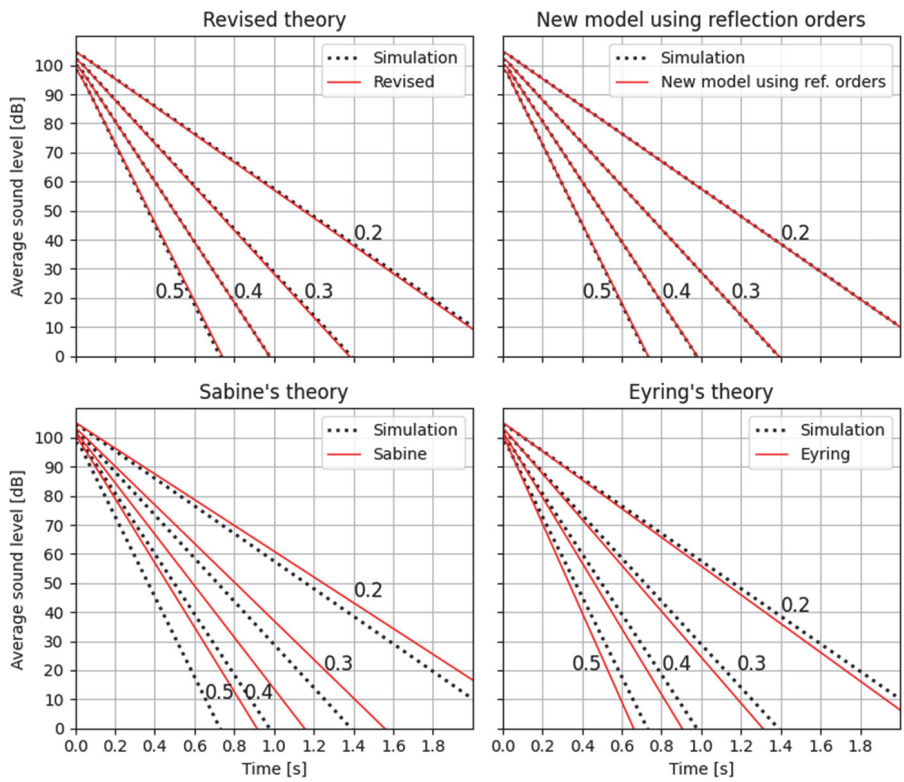

Cubic room with a side length of 10 m

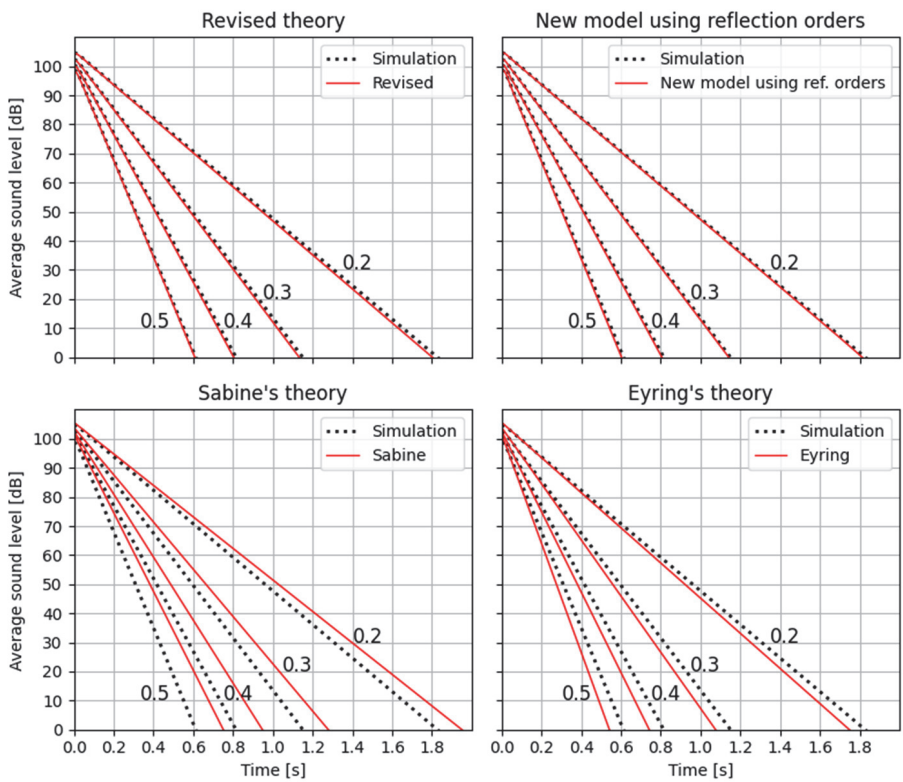

Rectangular room with dimensions of 15 × 10 × 5 m

**Fig. 6** Comparison of the theoretical reverberation decay curves with the simulation results.

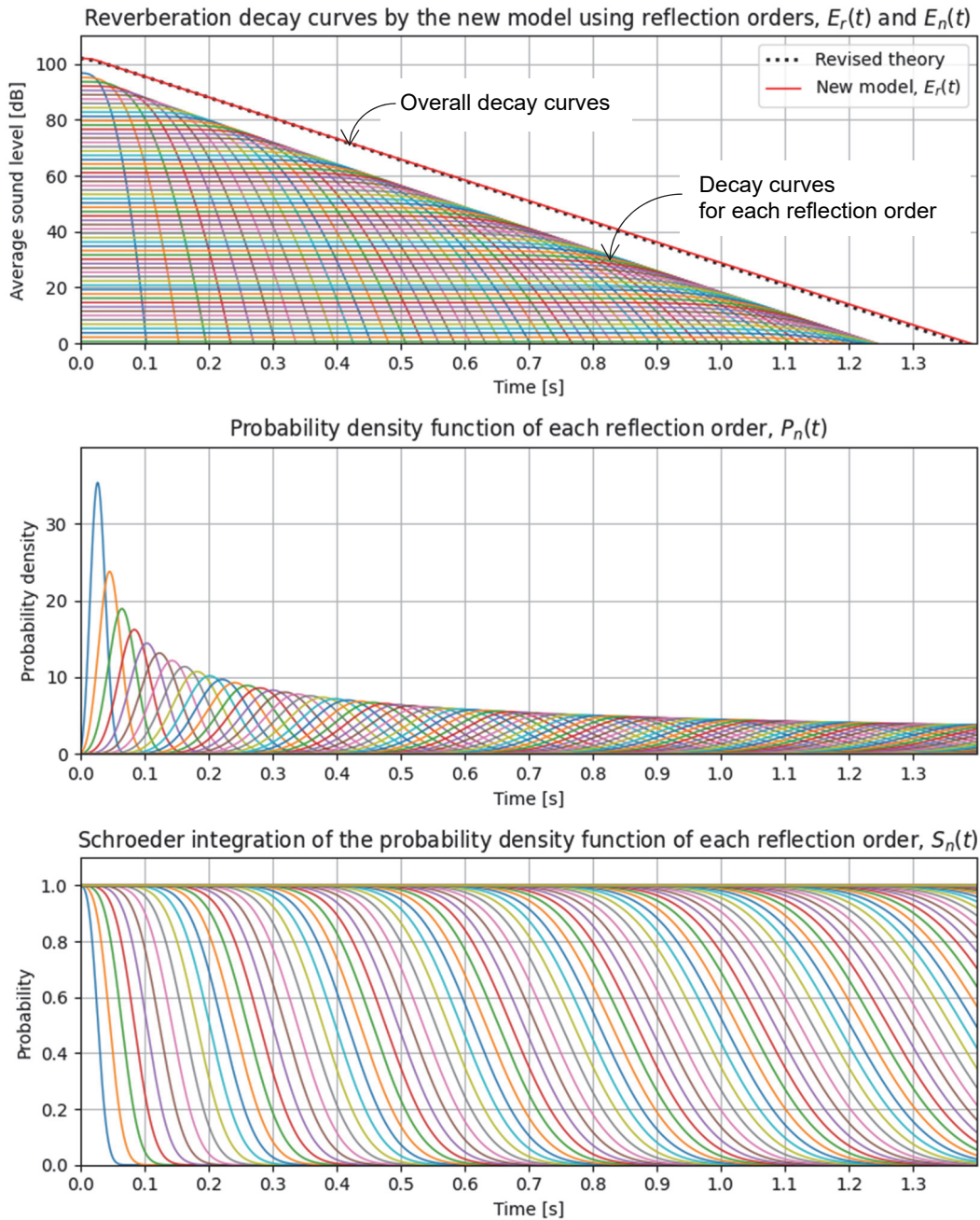

**Fig. 8** Results of the new model using reflection orders (Cubic room with a side length of 10 m, $\bar{\alpha} = 0.3$).

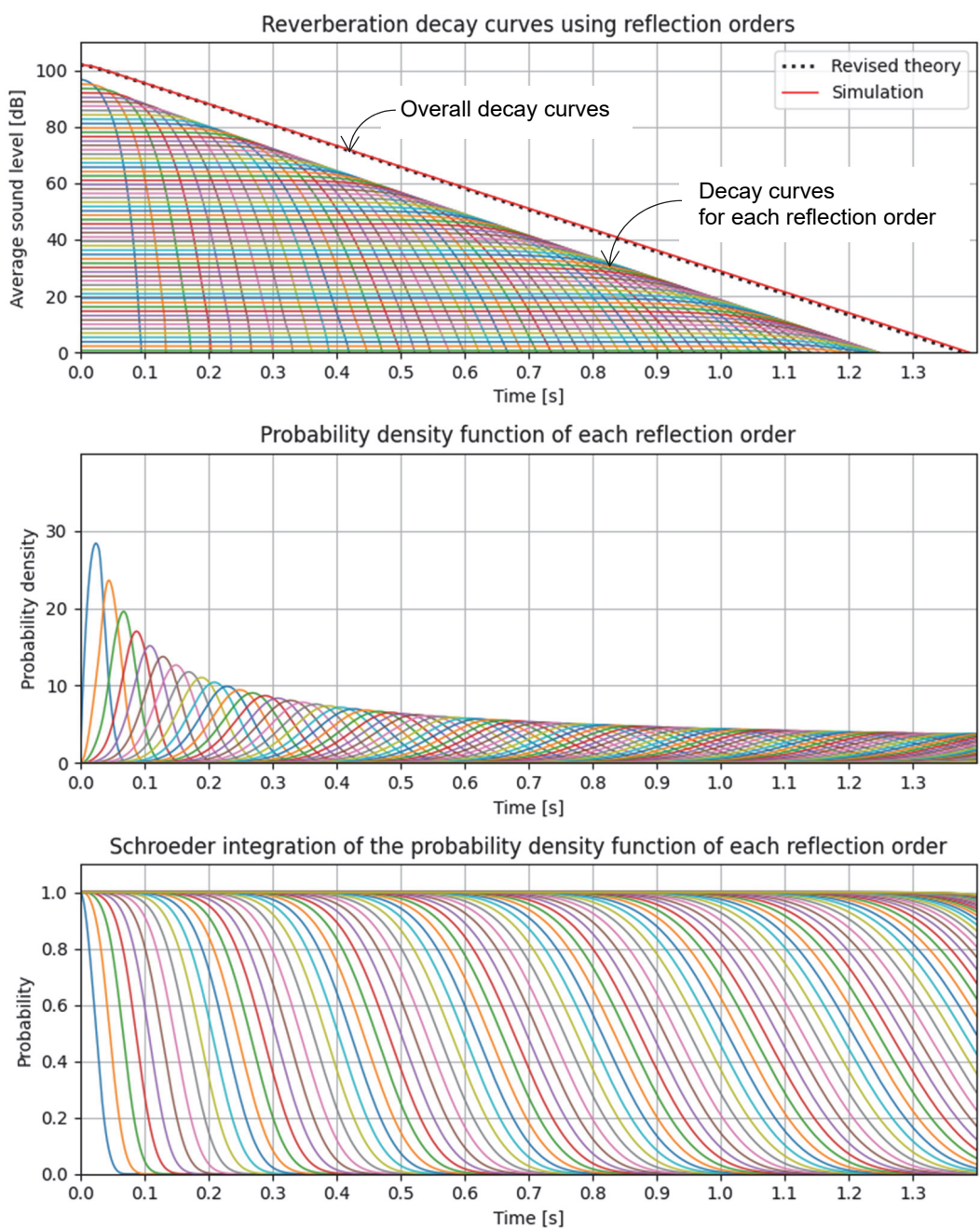

**Fig. 9** Results of the ray tracing simulation (Cubic room with a side length of 10 m, $\bar{\alpha} = 0.3$).

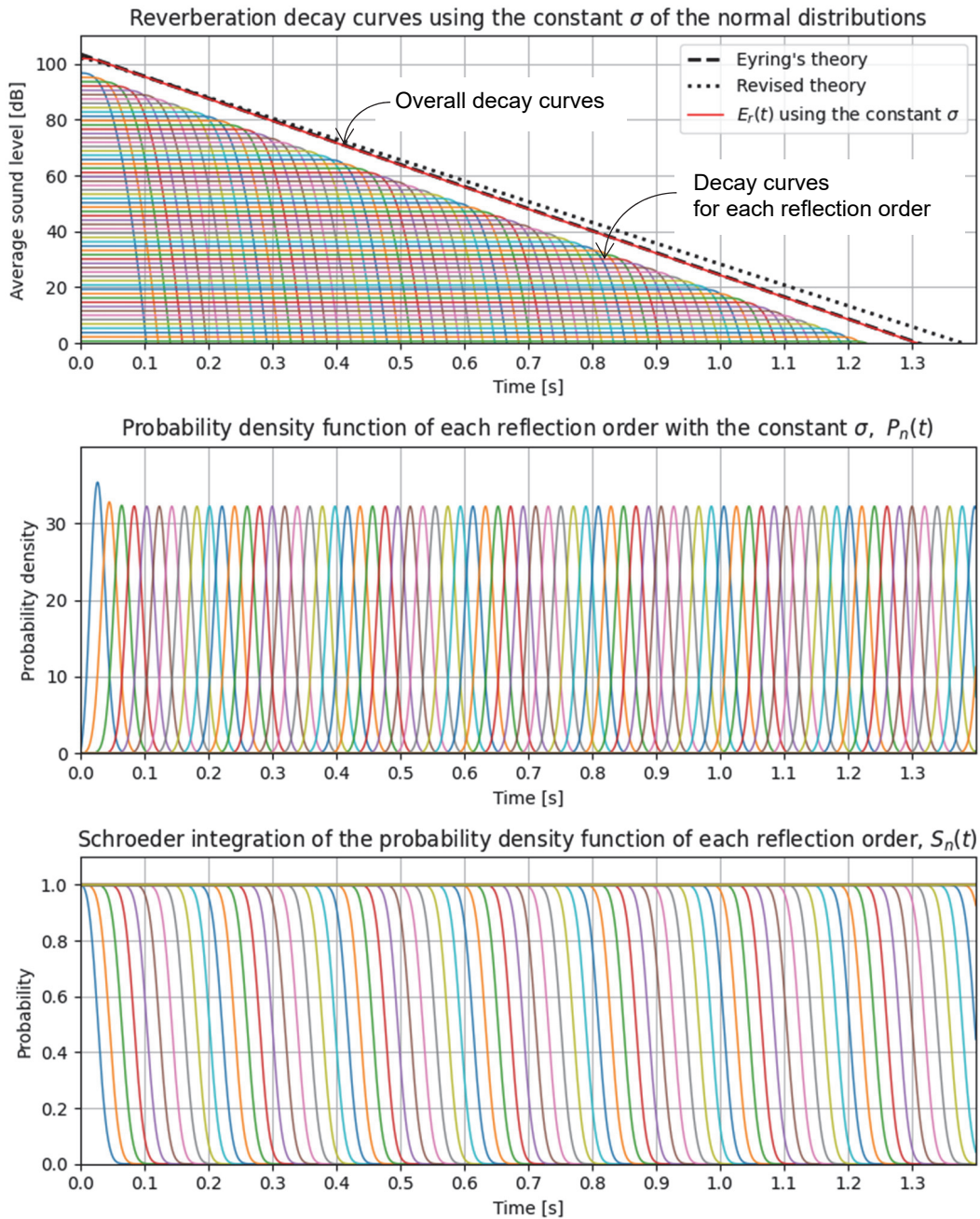

**Fig. 10** Interpretation of Eyring's theory in terms of the probability density function of each reflection order (Cubic room with a side length of 10 m, $\bar{\alpha} = 0.3$).